\documentclass[letterpaper, 10 pt, conference]{ieeeconf}
\IEEEoverridecommandlockouts 
\usepackage{cite}
\usepackage{amsmath,amssymb,amsfonts}
\usepackage{graphicx}
\usepackage{caption}
\usepackage{floatrow}
\usepackage{subcaption}
\usepackage{textcomp}
\usepackage{amsfonts, amsmath}
\usepackage{amssymb}
\usepackage{amsmath}
\usepackage{graphicx}
\usepackage{multirow}
\usepackage{etoolbox}
\makeatletter
\patchcmd{\@makecaption}
  {\scshape}
  {}
  {}
  {}
\makeatother
\usepackage[colorlinks=true,linkcolor=blue,citecolor=blue,urlcolor=blue]{hyperref}

\newtheorem{theorem}{Theorem}
\newtheorem{lemma}{Lemma}

\newtheorem{remark}{Remark}
\newtheorem{assumption}{Assumption}

\usepackage{cuted}
\usepackage[english]{babel}
\usepackage{xcolor}
\definecolor{hcolor}{rgb}{0.07,0.2,0.8}
    
    \makeatletter
 \let\old@ps@headings\ps@headings
 \let\old@ps@IEEEtitlepagestyle\ps@IEEEtitlepagestyle
 \def\confheader#1{%
 \def\ps@headings{%
 \old@ps@headings%
 \def\@oddhead{\strut\hfill#1\hfill\strut}%
 \def\@evenhead{\strut\hfill#1\hfill\strut}%
 }%
 \def\ps@IEEEtitlepagestyle{%
 \old@ps@IEEEtitlepagestyle%
 \def\@oddhead{\strut\hfill#1\hfill\strut}%
 \def\@evenhead{\strut\hfill#1\hfill\strut}%
 }%
 \ps@headings%
 }
 \makeatother

 \usepackage[pscoord]{eso-pic}

  \makeatletter
\newcommand{\linebreakand}{%
  \end{@IEEEauthorhalign}
  \hfill\mbox{}\par
  \mbox{}\hfill\begin{@IEEEauthorhalign}
}
\makeatother  

\begin{document}

\title{State and Input Constrained Model Reference Adaptive Control with Robustness and Feasibility Analysis
}

\author{Poulomee~Ghosh and Shubhendu~Bhasin
\thanks{Poulomee Ghosh and Shubhendu Bhasin are with Department of Electrical Engineering, Indian Institute of Technology Delhi, New Delhi, India. 
        {\tt\small (Email: poulomee.ghosh@ee.iitd.ac.in, sbhasin@ee.iitd.ac.in)}}}

\maketitle

\begin{abstract}We propose a model reference adaptive controller (MRAC) for uncertain linear time-invariant (LTI) plants with user-defined state and input constraints in the presence of unmatched bounded disturbances. Unlike popular optimization-based approaches for constrained control, such as model predictive control (MPC) and control barrier function (CBF) that solve a constrained optimization problem at each step using the system model, our approach is optimization-free and adaptive; it combines a saturated adaptive controller with a barrier Lyapunov function (BLF)-based design to ensure that the plant state and input always stay within pre-specified bounds despite the presence of unmatched disturbances. To the best of our knowledge, this is the first result that considers both state and input constraints for control of uncertain systems with disturbances and provides sufficient feasibility conditions to check for the existence of an admissible control policy. Simulation results, including a comparison with a robust MRAC, demonstrate the effectiveness of the proposed algorithm.

\end{abstract}


\section{Introduction}
\label{sec:intro}
 The design of constraint-compliant controllers is crucial for safety-critical systems, especially in the presence of plant uncertainty and disturbances. Classical adaptive control approaches, e.g., model reference adaptive controllers (MRAC), are primarily 
designed for systems with parametric uncertainty \cite{classMRAC, landau1984adaptive, mrac111}; however, under user-defined state and input constraints, they have limited practical applicability. The problem becomes more challenging in the presence of bounded external disturbances that can potentially destabilize the system and lead to constraint violation. Designing a robust MRAC algorithm that ensures reference model tracking while complying with state and input constraints, even in the presence of disturbances, is both practically relevant and theoretically challenging.\\
Various control strategies have been proposed to address the issue of output-constrained adaptive tracking, including model predictive control \cite{camacho2007constrained, zheng1995stability}, optimal control  \cite{opt1},\cite{opt}, invariant set theory \cite{blanchini, set}, reference governor \cite{rga}, \cite{gilbert}, control barrier function (CBF)-based approach \cite{CBF}, etc. However, most methods require computationally intensive optimization routines. Extensive work has also been done on tracking control of state-constrained nonlinear systems using barrier Lyapunov function (BLF)-based schemes that do not require an optimization step \cite{BLF, BLF2}. A disadvantage of using BLF-based strategies is that they often require significant control effort to push the state back towards the safe set whenever it approaches the safety boundary. This may lead to actuator failure and compromise system reliability.\\
Several approaches have also been proposed to tackle the problem of input constraints or actuator saturation, especially using various saturated functions, e.g., hyperbolic tangent, sigmoid, etc. \cite{incon10, inconnew, incon11, incon12, incon13}. 
Anti-windup strategies have also been used to address the problem of systems with input constraints \cite{annaswamy}, \cite{Lav}. \\
 Most constrained control approaches typically focus on either state or input constraints; however, few control strategies efficiently manage both, especially for uncertain plants. MPC \cite{dhar, mpc11, dhar2} integrates state and input constraints within the optimization routine, albeit with increased computational complexity. Moreover, limited or imperfect model knowledge often yields overly conservative results due to robust constraint tightening in robust and adaptive variants of MPC. The analysis of recursive feasibility and stability of MPC design is further complicated in the presence of system uncertainty.
A few recent works \cite{Anderson, chattopadhyay2025model1, ppc} develop adaptive laws that impose user-defined bounds on both the tracking error and the input. The method in \cite{ppc} achieves prescribed performance while satisfying the input constraint by dynamically modifying the reference trajectory in response to saturation. These approaches, while effective, rely on reference model modification, which may not be suitable for scenarios where altering the reference signal is undesirable.

Due to the theoretical challenges associated with handling both state and input constraints, few results exist that address this practical problem for uncertain systems. In our previous works \cite{mymrac, myel}, we designed an adaptive controller using a BLF-based design combined with a saturated controller to deal with both input and state constraints; however, the results lacked analysis of both feasibility and robustness.\\
In this work, we propose a control architecture that provides sufficient feasibility conditions for the state, input, and disturbance bounds. We develop a BLF-based saturated MRAC architecture for uncertain multivariable LTI systems in the presence of bounded external disturbances while adhering to user-defined state and input constraints. The control law is made robust to external disturbances by using a projection operator, yielding a locally stable result. No additional conditions are imposed on the reference model or the reference input, making this method a general tool for a wide range of applications.\\
A pertinent problem in constrained control is the existence of a feasible control policy. While it is clear that the bounds on state, input, and disturbance are related and, therefore, cannot be chosen arbitrarily and independently, it is not apparent how an analytical expression that manifests this interrelationship and the associated trade-offs might be obtained. A major contribution of this work is to provide verifiable feasibility conditions to check for the existence of a control law that ensures compliance with user-defined state and input constraints in the presence of external disturbances. \\
The primary contributions of this work are summarized as follows.
\begin{enumerate}
\item We design an adaptive controller that effectively manages both the state and input constraints in the presence of parametric uncertainties and unmatched disturbances in the plant model. 


\item We present verifiable feasibility conditions for the state, input, and disturbance bounds. Further, we highlight the trade-off between the state constraint, the input constraint, and the disturbance bound, thus providing insights into how these constraints are coupled and their effect on the system performance.
\end{enumerate}

Throughout this paper, we use $\mathbb{R}$ to represent the set of real numbers, $\mathbb{R}^{p \times q}$ to denote the set of $p\times q$ real matrices and $I_{p}$ to denote the identity matrix of ${p \times p}$ dimension. $\log(\cdot)$ denotes the natural logarithm of $(\cdot)$, and the Euclidean vector norm and corresponding equi-induced matrix norm are represented by $\|\cdot\|$. The trace of $A\in \mathbb{R}^{n \times n}$ is represented by $\text{tr}(A)$ and the eigenvalues of A with minimum and maximum real parts are denoted by $\lambda_{min}\{A\}$ and $\lambda_{max}\{A\}$, respectively.


\section{Problem Formulation}
\label{sec:LFSR}

Consider a multivariable linear time-invariant (LTI) plant in the presence of unmatched bounded external disturbances.  
\begin{align}
    \dot{x}(t)=Ax(t)+Bu(t)+d(t)
    \label{plant}
\end{align}
where $x(t)\in \mathbb{R}^n$ is the system state and $u(t) \in \mathbb{R}^m$ is the input, $A \in \mathbb{R}^{n \times n}$ and $B \in \mathbb{R}^{n \times m}$ are the system and input matrices, respectively. It is assumed that $n\geq m$. Here, $A$ is unknown while $B$ is known and has full column rank, and the pair \((A,B)\) is stabilizable. The unmatched external disturbance $d(t) \in \mathbb{R}^n$ is bounded, i.e. $\|d(t)\|\leq \bar{d}$, where $\bar{d}$ is a known positive constant. \\
A stable known reference model is considered
\begin{align}
    \dot{x}_r(t)=A_rx_r(t)+B_rr(t)
    \label{ref}
\end{align}
where $x_r(t)\in \mathbb{R}^n$ denotes the reference model state and $\|x_r(t)\|<\bar{\mathcal{X}}_r$, where $\bar{\mathcal{X}}_r$ is a known positive constant. $r(t) \in \mathbb{R}^{m}$ represents a piecewise continuous bounded reference input such that $\|r(t)\|\leq \bar{r}$, where $\bar{r}$ is a known positive constant. $A_r \in \mathbb{R}^{n \times n}$ and $B_r \in \mathbb{R}^{n \times m}$ are known state and input matrices, respectively. $A_r$ is assumed to be Hurwitz, i.e., for every symmetric and positive definite matrix  $Q\in\mathbb{R}^{n\times n}$, there exists a symmetric and positive definite matrix $P\in \mathbb{R}^{n\times n}$, such that $A_r$ satisfies 
\begin{align}
  A_r^TP+PA_r+Q=0 
  \label{hurwitz}
\end{align}
The trajectory tracking error dynamics is given by
\begin{align}
   e(t) \triangleq x(t)-x_r(t) 
\end{align}
The control objective is to design a control input $u(t)$ such that $x(t)$ tracks $x_r(t)$ while satisfying the following state and input constraints.\\
\textbf{State Constraint:}
The plant states are uniformly bounded, i.e. $\|x(t)\|< \bar{\mathcal{X}}$ $\forall t \geq 0$, where $\bar{\mathcal{X}}$ is a user-defined positive constant.\\
\textbf{Input Constraint:} The amplitude of the control input is uniformly bounded, i.e. $\|u(t)\|\leq \bar{\mathcal{U}}$ $\forall t \geq 0$, where $\bar{\mathcal{U}}$ is a user-defined positive constant.


\section{Proposed Methodology}

\subsection{Saturated Controller Design}

Consider the LTI plant (\ref{plant}) and reference model (\ref{ref}). To account for actuator saturation, 
an auxiliary control input $v(t)\triangleq[v_1(t),\hdots, v_m(t)]^T \in \mathbb{R}^m$ is designed as
\begin{align}
&v(t)=\hat{K}_x(t)x(t)+\hat{K}_r(t)r(t)
\label{pc1}
\end{align}
where $\hat{K}_x(t)\in \mathbb{R}^{m\times n}$ and $\hat{K}_r(t)\in \mathbb{R}^{m\times m}$ are the estimates of the true controller parameters $K_x\in \mathbb{R}^{m\times n}$ and $K_r\in \mathbb{R}^{m\times m}$, respectively.
\begin{assumption}
\label{assumption_matching_condition}
    There exist controller parameters $K_x$ and $K_r$ such that the following matching conditions are satisfied.
    \begin{align}
        &A+BK_x=A_r \label{mc1} \\
        &BK_r=B_r \label{mc2}
    \end{align}
\end{assumption}
The above assumption is standard in adaptive control literature \cite{classMRAC}. 
\begin{assumption}
\label{assumption_kxbar}
The norm of controller parameters is bounded such that $\|K_x\|\leq \bar{K}_x$ and $\|K_r\|\leq \bar{K}_r$, where $\bar{K}_x, \bar{K}_r>0$ are assumed to be known.
\end{assumption}
\begin{remark}
    Assumption \ref{assumption_kxbar} is standard in projection-based adaptive control literature \cite{lavretsky2012robust}. One way to estimate the bound on $\|K_x\|$ is to use the knowledge of the upper bound of the matrix $A$, if available. From matching condition (\ref{mc1}), we can obtain

    \begin{align}
    K_x=B^{\dagger}(A_r-A)
        \label{abar0}
    \end{align}
    where ${B}^{\dagger}\in \mathbb{R}^{m \times n}$ represents the left inverse of the matrix $B \in \mathbb{R}^{n \times m}$ given by ${B}^{\dagger}=(B^TB)^{-1}B^T$, whose existence is ensured since $B$ has full column rank.  
    We can derive $\bar{K}_x$ from (\ref{abar0}), as shown below.
    \begin{align}
        \bar{K}_x\geq \|{B}^{\dagger}\|\|{A}_r-{A}\|
        \label{abar}
    \end{align}
\end{remark}
To ensure that the control input remains within a specified limit, a saturated feedback controller is designed as
\begin{align}
&u_i(t)=\begin{cases}
v_i(t) & \text{if } \|v(t)\| \leq \bar{\mathcal{U}} \\
v_i(t) \cdot \frac{\bar{\mathcal{U}}}{\|v(t)\|} & \text{if } \|v(t)\| > \bar{\mathcal{U}}
\end{cases},
\hspace{4pt} i=1, \hdots , m
\label{pc2}
\end{align}
where $u(t)\triangleq[u_1(t),\hdots, u_m(t)]^T $. The closed-loop trajectory tracking error with input saturation is given as
\begin{align}
\dot{e}(t)=&A_re(t)+B\tilde{K}_x(t)x(t)+B\tilde{K}_r(t)r(t)+B\Delta u(t)\nonumber \\
&+d(t)
    \label{edot}
\end{align}
where $\tilde{K}_x(t)\triangleq \hat{K_x}(t)- K_x\in\mathbb{R}^{m \times n}$ and $\tilde{K}_r(t)\triangleq \hat{K_r}(t) - K_r\in \mathbb{R}^{m \times m}$ are the parameter estimation errors and the difference between actual control input and auxiliary control input is defined by 
\begin{align}
    \Delta u(t) \triangleq u(t)-v(t)
\end{align}
To transform the state constraint into the constraint on tracking error, we introduce an auxiliary reference model defined as
\begin{align}
    x_a(t) =
\begin{cases}
x_r(t) & \text{if } \|x_r(t)\| < \bar{\mathcal{X}} \\
{x_r(t)}\cdot\frac{\bar{\mathcal{X}}_a} {\|x_r(t)\|} & \text{otherwise}\\
\end{cases}
    \label{auxref}
\end{align}
where, $x_a(t)\in \mathbb{R}^n$ denotes auxiliary reference model states, bounded such that $\|x_a(t)\|\leq\bar{\mathcal{X}}_a$, where $\bar{\mathcal{X}}_a>0$ is a known constant, chosen such that the auxiliary reference states remain within the user-defined safe set, i.e. $\bar{\mathcal{X}}_a < \bar{\mathcal{X}}$ $\forall t \geq 0$.
\begin{remark}
    By introducing an auxiliary reference model, we can relax the common assumption, typically used in most BLF-based approaches, that the actual reference model states must always lie within the safe set. Note that, using (\ref{auxref}),  the actual reference trajectory remains unmodified, preserving the original design objectives without imposing additional restrictions. This relaxation provides flexibility in designing the reference trajectory while ensuring that the plant states and the trajectory tracking error always remain within the safe region. Although this flexibility may come at the cost of tracking performance,  this trade-off is justified as guaranteeing constraint satisfaction is our primary objective.
\end{remark}
Provided (\ref{auxref}), the constraint on plant state can be transformed to the constraint on the trajectory tracking error: $\|e(t)\|<\xi$, $\forall t\geq 0$, where $\xi\triangleq\bar{\mathcal{X}}-\bar{\mathcal{X}}_a>0$ is a positive constant. Clearly,  $\|e(t)\|<\xi$ implies $\|x(t)\|<\bar{\mathcal{X}}$, $\forall t \geq 0$.

\subsection{BLF-based design for state constraint satisfaction} \label{AA}
In this section, we introduce a BLF-based design of the adaptive update laws \cite{BLF} to satisfy constraints on plant state and guarantee boundedness of all the closed-loop signals.

\begin{lemma}
Consider two open sets defined as $\Omega_e := \{e \in \mathbb{R}^n : \|e\|<\xi \}\subset \mathbb{R}^n$ and $\Psi:=\Omega_e \times \mathbb{R}^N  \subset \mathbb{R}^{n+N}$. Let the system dynamics of $\mu(t):=[e(t),\zeta(t)]^T\in \Psi$ is given by
\begin{align}
    \dot{\mu}=f(t,\mu)
\end{align}
where $e(t)$ is the constrained state, $\zeta(t):=[\tilde{K}_x(t),\tilde{K}_r(t)]$ denotes the unconstrained states, and the function $f:\mathbb{R}_{+}\times \Psi \rightarrow \mathbb{R}^{N+n}$ is measurable and locally Lipschitz in $\mu$, piecewise continuous and locally integrable on $t$. Let there exist continuously differentiable and positive definite functions $V_1(e):\Omega_e \rightarrow \mathbb{R}_{+}$ and $V_2(\zeta):\mathbb{R}^N\rightarrow\mathbb{R}_{+}$, such that
\begin{align}
    &V_1(e)\rightarrow \infty \hspace{18pt} \text{as} \hspace{4pt} \|e\|\rightarrow \xi \\
    &\alpha_1(\|\zeta\|)\leq V_2(\zeta)\leq \alpha_2(\|\zeta\|)
\end{align}
where $\alpha_1$ and $\alpha_2$ are class $\mathcal{K}_{\infty}$ functions. Let $V(\mu)=V_1(e)+V_2(\zeta)$. Provided $e(0)\in \Omega_e$, if the following inequality holds,
\begin{align}
   \dot{V}< 0
    \label{uub0}
\end{align}
then $e(t)\in \Omega_e$ and $\zeta(t)$ is bounded $\forall t >0$.\\
\begin{proof}
From \cite{sontag2013mathematical}, the existence of a unique maximal solution $\mu(t)$ can be proved for a time interval $[0,t_m)$ that ensures the existence of $V(\mu(0))$.  
Now, integrating (\ref{uub0}), it can be shown that
\begin{align}
   V(\mu(t))<V(\mu(0)) &&\forall t\in [0,t_m)
\end{align}
As $V(\mu(t))$ is bounded, $V_1(e)$ also remains bounded $\forall t\in [0,t_m)$ which implies that $\zeta(t)$ is bounded and $\|e(t)\|<\xi$, $\forall t\in [0,t_m)$. Consequently, it can be shown that $\mu(t)$ exists $\forall t\in [0,\infty)$, which proves $e(t)\in \Omega_e$ $\forall t\in [0,\infty)$.
\end{proof}
\end{lemma}
Since the state constraint is transformed to the constraint on tracking error, consider a BLF defined on the set $\Omega_e^{'}:\{e\in\mathbb{R}^n: e^TPe < \xi^{'^2}\}$.

\begin{align}
    &V_1(e) \triangleq \frac{1}{2} \log{ \frac{\xi^{'2}}{\xi^{'2}-e^TPe}}
\end{align}
where, $\xi^{'}=\xi\sqrt{\lambda_{min}\{P\}}$. As the state $e(t)$ approaches the boundary of the set $\Omega_e$, the BLF $V_1(e)\rightarrow \infty$.
For the proposed controller (\ref{pc1}) and (\ref{pc2}), the following adaptive update laws are defined based on the subsequent Lyapunov analysis.
\begin{subequations}
\begin{align}
     &\dot{\hat{K}}_x=\text{proj}_{\Omega_1}\bigg(-\frac{\Gamma_xB^TPex^T}{\xi^{'^2}-e^TPe}\bigg)\\
    &\dot{\hat{K}}_r=\text{proj}_{\Omega_2}\bigg(-\frac{\Gamma_r B^TPer^T}{\xi^{'^2}-e^TPe} \bigg)
\end{align}
\label{proposedlaw}
\end{subequations}
where $\Gamma_x,\Gamma_r\in \mathbb{R}^{m\times m}$ are positive definite adaptive gain matrices and $P\in \mathbb{R}^{n\times n}$ is defined in (\ref{hurwitz}). The projection operator \(\text{proj}_{*}(\cdot)\) \cite{lavretsky2012robust} ensures that the parameter update law \((\cdot)\) keeps the parameters bounded within a convex and compact region, denoted by `\(*\)', in the parameter space. In this context, the convex functions associated with the projection operator are chosen as \(f(\hat{K}_x) = \|\hat{K}_x\|^2\), \(f(\hat{K}_r) = \|\hat{K}_r\|^2\) and the convex and compact regions are defined by $\Omega_1=\{\hat{K}_x\in\mathbb{R}^{m\times n}|\|\hat{K}_x\|^2 \leq \bar{K}_x^2\}$ and \(\Omega_2=\{\hat{K}_r\in\mathbb{R}^{m \times m}|\|\hat{K}_r\|^2 \leq \bar{K}_r^2\}\), which align with Assumption \ref{assumption_kxbar}. Furthermore, by invoking \cite[Lemma~6]{lavretsky2011projection}, it can be asserted that the projection operator introduces a negative semidefinite term in the Lyapunov derivative when the estimate is at the boundary of the convex and compact region, thereby preserving the stability results.

\begin{theorem}
For the plant and reference model in (\ref{plant}) and (\ref{ref}), respectively, provided the initial tracking error $e(0)\in \Omega_e$, Assumptions \ref{assumption_matching_condition}-\ref{assumption_kxbar} hold, and the following feasibility condition (C1) is satisfied,
\begin{enumerate}
\item[\textbf{C1:}]The input  constraint satisfies the following inequality:
        \label{c1}
\begin{align}
   \bar{\mathcal{U}}> \bar{\mathcal{X}}(\bar{K}_x-\eta)+\eta\bar{\mathcal{X}}_a+\bar{K}_r\bar{r}+\frac{\bar{d}}{\|B\|}
    \label{c2}
\end{align}
where $\eta=\frac{\lambda_{min}\{Q\}}{2\lambda_{max}\{P\}\|B\|}>0$ is a known constant.
\end{enumerate}
the proposed controller (\ref{pc1}), (\ref{pc2}) along with the adaptive laws (\ref{proposedlaw}) ensure the following (P1-P3).

\begin{enumerate}
    \item [\textbf{P1:}] State constraint satisfaction, i.e.,$\|x(t)\|< \bar{\mathcal{X}}$ $\forall t \geq 0$.
    \item [\textbf{P2:}]  Input constraint satisfaction, i.e.,$ \|u(t)\|\leq \bar{\mathcal{U}}$ $\forall t \geq 0$.
    \item [\textbf{P3:}] The trajectory tracking error $\|e(t)\|$ is UUB, and all the closed-loop signals are bounded.
\end{enumerate}

\end{theorem}

\begin{proof}
Consider the candidate Lyapunov function $V(e,\tilde{K}_x,\tilde{K}_r):\Omega_e^{'}\times \mathbb{R}^N\rightarrow \mathbb{R}_{+}$ as,
\begin{align}
    V&=
    \frac{1}{2}
    \bigg[\log\frac{\xi^{'^2}}{\xi^{'^2}-e^TPe}+tr(\tilde{K}_x^T\Gamma_x^{-1}\tilde{K}_x)\nonumber\\
&+tr(\tilde{K}_r^T\Gamma_r^{-1}\tilde{K}_r)\bigg]
    \label{lyap}
\end{align}
Taking the time derivative of $V$ along the system trajectory 
\begin{align}
    \dot{V}=&\frac{1}{2(\xi^{'^2}-e^TPe)}\bigg[e^TP(A_re+B\tilde{K}_xx+B\tilde{K}_rr+B\Delta u \nonumber\\
    &+{d})+(A_re+B\tilde{K}_xx+B\tilde{K}_rr+B\Delta u+{d})^TPe\bigg] \nonumber\\
    &+tr(\tilde{K}_x^T\Gamma_x^{-1}\dot{\hat{K}}_x) +tr(\tilde{K}_r^T\Gamma_r^{-1}\dot{\hat{K}}_r) 
    \label{vdot}
    \end{align}
Substituting the adaptive update laws (\ref{proposedlaw}), 
\begin{align}
    \dot{V}\leq & -\frac{e^TQe}{2(\xi^{'^2}-e^TPe)}+\frac{e^TPB\Delta u}{\xi^{'^2}-e^TPe}+\frac{e^TPd}{\xi^{'^2}-e^TPe}
    \nonumber\\
      \leq & -\frac{\lambda_{min}\{Q\}\|e\|^2}{2(\xi^{'^2}-e^TPe)}+\frac{\lambda_{max}\{P\}\|e\|\|d\|}{\xi^{'^2}-e^TPe}
     \nonumber\\
    &+\frac{\lambda_{max}\{P\}\|B\|\|e\|\|\Delta u\|}{\xi^{'^2}-e^TPe}
    \label{lyapfunc0point5}
\end{align}

We now consider two cases depending on whether the controller is operating within the saturation limits (\textit{Case 1}) or is saturated (\textit{Case 2}). 
\\~\\
\textit{Case 1.1:} $\|v(t)\|\leq \bar{\mathcal{U}}$\\
    In this case, $u_i(t)=v_i(t)$ and $\Delta u_i(t)=0$, $\forall t \geq 0$, which implies $\|\Delta u(t)\|=0$ $\forall t \geq 0$.\\~\\
    \textit{Case 1.2:} $\|v(t)\|> \bar{\mathcal{U}}$\\
    In this case, $u_i(t)=v_i(t) \cdot \frac{\bar{\mathcal{U}}}{\|v(t)\|}$ and $\Delta u_i(t)=v_i(t) \cdot \frac{\bar{\mathcal{U}}}{\|v(t)\|}-v_i(t)$, which implies 
    \begin{align}
        \|\Delta u(t)\| & = \|v(t)\|(1-\frac{\bar{\mathcal{U}}}{\|v(t)\|})\\
                          & \leq
    \bar {K}_x(\|e\|+\|x_a\|)+ \bar{K}_r\bar{r}-\bar{\mathcal{U}}
    \end{align}
We proceed with the Lyapunov analysis using Case $1.2$, as Case $1.1$ is subsumed in the more general Case $1.2$.
Substituting the bound on $\|\Delta u(t)\|$ in ($\ref{lyapfunc0point5}$),
\begin{align}
     \dot{V}\leq & -\frac{\lambda_{min}\{Q\}\|e\|^2}{2(\xi^{'^2}-e^TPe)}+\frac{\lambda_{max}\{P\}\|e\|\|d\|}{\xi^{'^2}-e^TPe}\nonumber\\
    &+\frac{\lambda_{max}\{P\}\|B\|\|e\|}{\xi^{'^2}-e^TPe}[\bar {K}_x(\|e\|+\|x_a\|) +\bar{K}_r\bar{r}-\bar{\mathcal{U}}]
    \nonumber\\
   \leq & -\frac{\lambda_{min}\{Q\}\|e\|}{2(\xi^{'^2}-e^TPe)}\bigg[\frac{1}{\eta}\varrho
   -\sigma\|e\|\bigg]
    \label{lyapfunc2}
\end{align}
where $\varrho=\bar{\mathcal{U}}-\bar{K}_x\bar{\mathcal{X}}_a-\bar{K}_r\bar{r}-\frac{\bar{d}}{\|B\|}$, $\sigma=\frac{\bar{K}_x}{\eta}-1$ and $\eta=\frac{\lambda_{min}\{Q\}}{2\lambda_{max}\{P\}\|B\|}$ are known constants. 
Based on the sign of $\sigma$, we will now consider two cases.\\
\textit{Case 2.1 ($\sigma>0$):}
Since $e(0)\in \Omega_e$, which implies $(\xi^{'^2}-e(0)^TPe(0))>0$, at $t=0$, to achieve a stable result from (\ref{lyapfunc2}), the following condition must hold.
\begin{align}
    \xi<\frac{\bar{\mathcal{U}}-\bar{K}_x\bar{\mathcal{X}}_a-\bar{K}_r\bar{r}-\frac{\bar{d}}{\|B\|}}{\bar{K}_x-\eta}
    \label{con1}
\end{align}
which implies feasibility condition C1 since \(\xi=\bar{\mathcal{X}}-\bar{\mathcal{X}}_a\). Note that satisfying C1 also implies $\varrho>0$, which is a necessary condition for feasibility in this case. Employing Lemma 1, at $t=0$, (\ref{lyapfunc2}) can be written as
\begin{align}
    \dot{V}(0)< 0
    \label{uub}
\end{align}
\textit{Case 2.2 $(\sigma<0)$:} Following a similar argument as case 2.1, to achieve a stable result from (\ref{lyapfunc2}), we are considering two scenarios. If $\varrho\geq0$ we can prove (\ref{uub}) without any other feasibility condition. On the other hand, if $\varrho<0$, (\ref{uub}) can be proved from (\ref{lyapfunc2}) if the feasibility condition C1 holds.

Using Lemma 1, it can be proved from (\ref{uub}) that $e(0^+)\in \Omega_e$ at $t=0^+$, provided $e(0)\in \Omega_e$ and (\ref{con1}) is satisfied. Repeating the above argument for all time instants,
\begin{align}
    \dot{V}(t)< 0 && \forall t \geq 0
    \label{uub1}
\end{align}
For any $P>0$, $e^TPe\geq \lambda_{min}\{P\}\|e\|^2$ and it can be proved from (\ref{uub1}) that $\|e(t)\|< \xi$ $\forall t \geq 0$, i.e., the trajectory tracking error remains bounded within the pre-specified bound: $e(t) \in \Omega_e$ $\forall t\geq0$. \\
Further, since $x(t)=e(t)+x_a(t)$, $\|x_a(t)\|\leq \bar{\mathcal{X}}_a$ and $\|e(t)\|<\xi$, the proposed control policy guarantees that the plant states remain confined within the user-defined safe limit.
\begin{align}
    \|x(t)\|< \xi+\bar{\mathcal{X}}_a= \bar{\mathcal{X}} && \forall t \geq 0
\end{align}
Therefore, the state constraint is satisfied, i.e., $\|x(t)\|< \bar{\mathcal{X}}$ for all $t\geq 0$.
Since $e(t)$, $\tilde{K}_x(t)$, $\tilde{K}_r(t)$ $\in \mathcal{L}_{\infty}$, and $K_x$ and $K_r$ are constants, it can be proved that $\hat{K}_x(t), \hat{K}_r(t) \in \mathcal{L}_{\infty}$ which further guarantees that the plant state $x(t)$ and control input $u(t)$ remain bounded for all time. Thus, the proposed controller ensures that all the closed-loop signals are bounded.
\end{proof}

\section{Feasibility Analysis} \label{feas}
\begin{figure}[h!]
     \centering
     \begin{subfigure}[b]{0.4\textwidth}
         \centering
         \includegraphics[width=\textwidth]{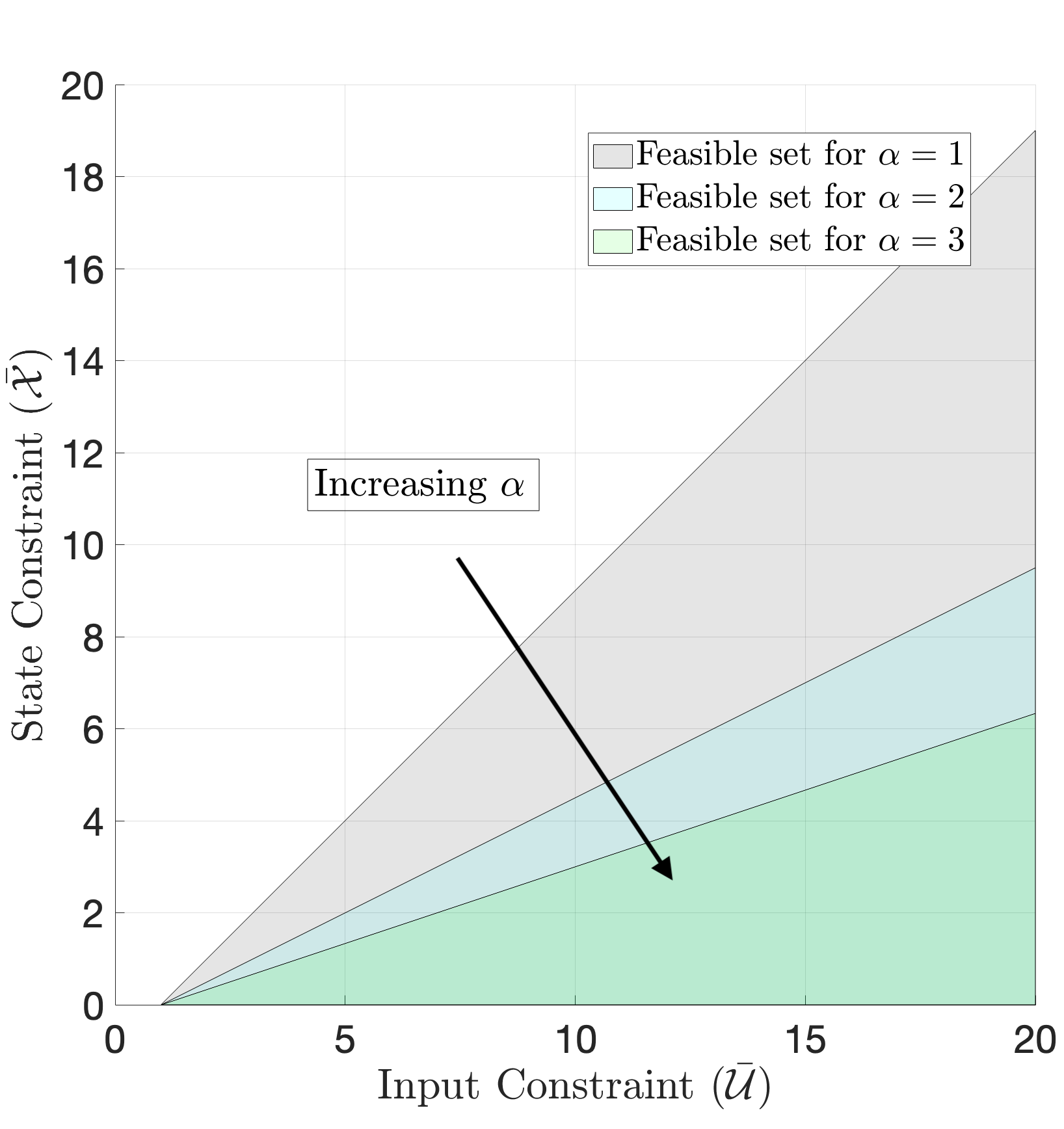}
         \caption{}
         \label{feasible_alpha}
     \end{subfigure}
     ~
     \begin{subfigure}[b]{0.4\textwidth}
         \centering
         \includegraphics[width=\textwidth]{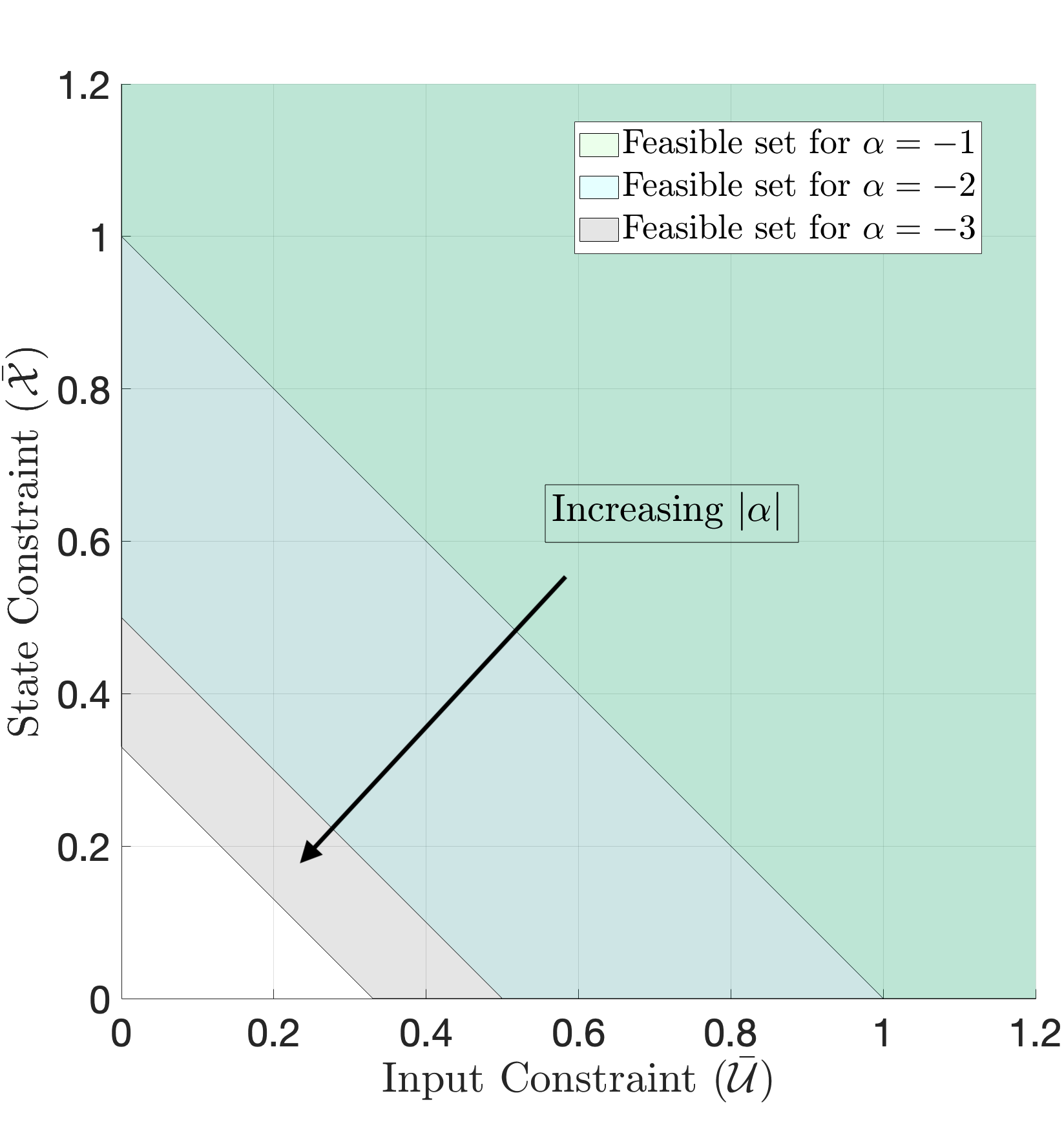}
         \caption{}
         \label{feasible_negalpha}
     \end{subfigure}
     \hfill
     \begin{subfigure}[b]{0.4\textwidth}
         \centering
         \includegraphics[width=\textwidth]{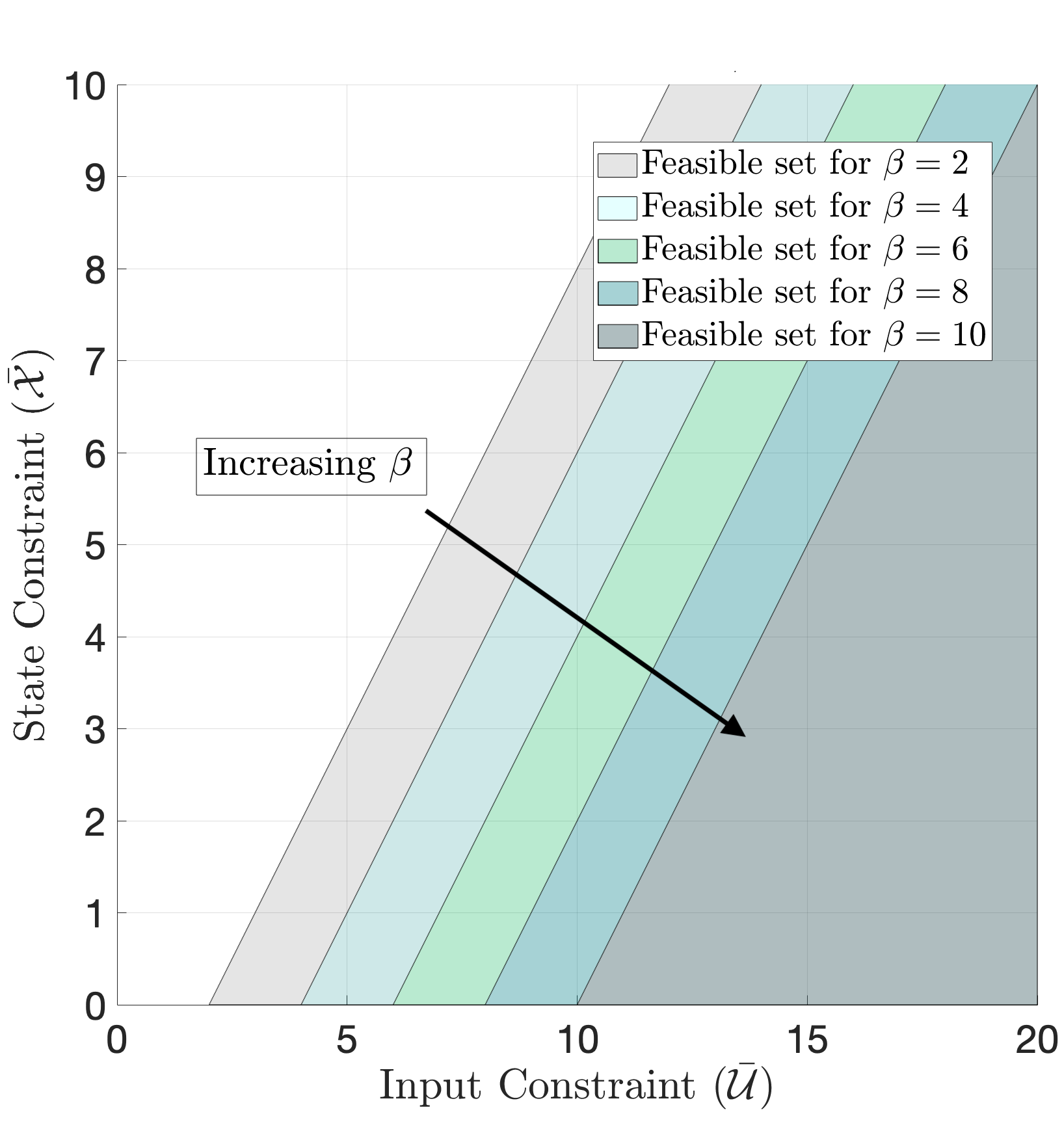}
         \caption{}
         \label{Hardcon_both}
     \end{subfigure}
     ~
     \begin{subfigure}[b]{0.4\textwidth}
         \centering
         \includegraphics[width=\textwidth]{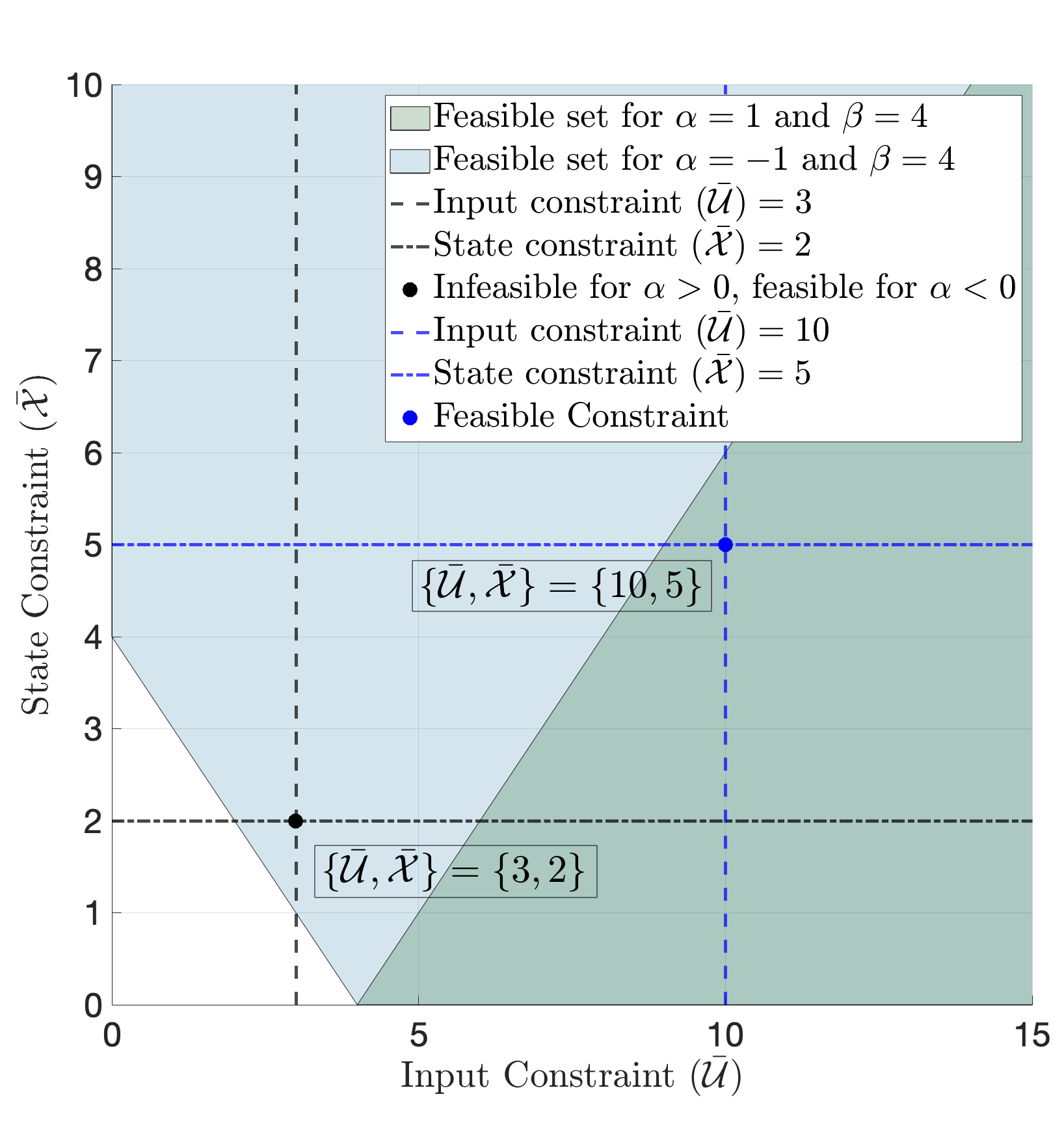}
         \caption{}
         \label{Hardcon_state}
     \end{subfigure}
     \hfill
     \begin{subfigure}[b]{0.4\textwidth}
         \centering
         \includegraphics[width=\textwidth]{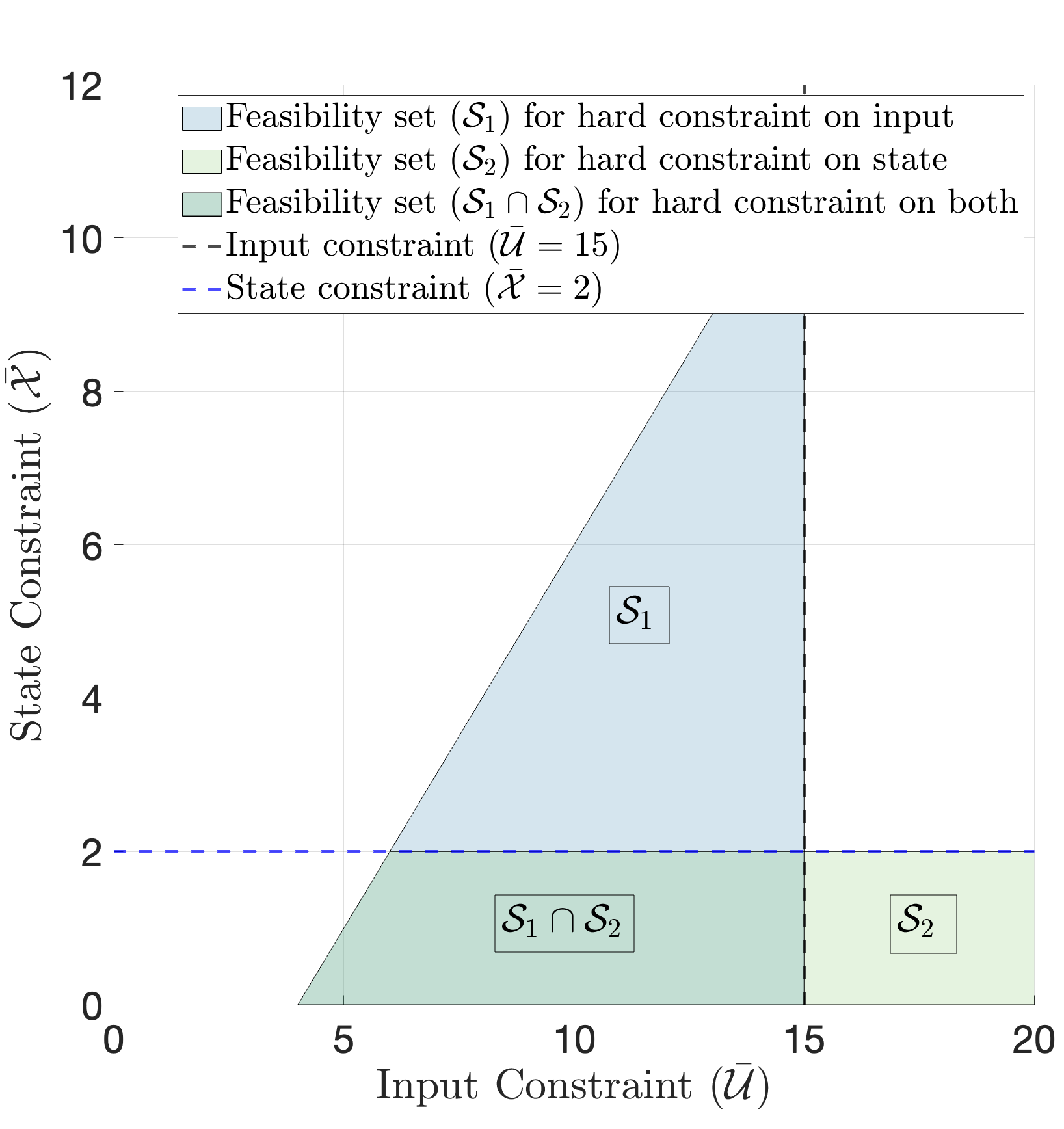}
         \caption{}
         \label{hardcon_input}
     \end{subfigure}
     ~
     \begin{subfigure}[b]{0.4\textwidth}
         \centering
         \includegraphics[width=\textwidth]{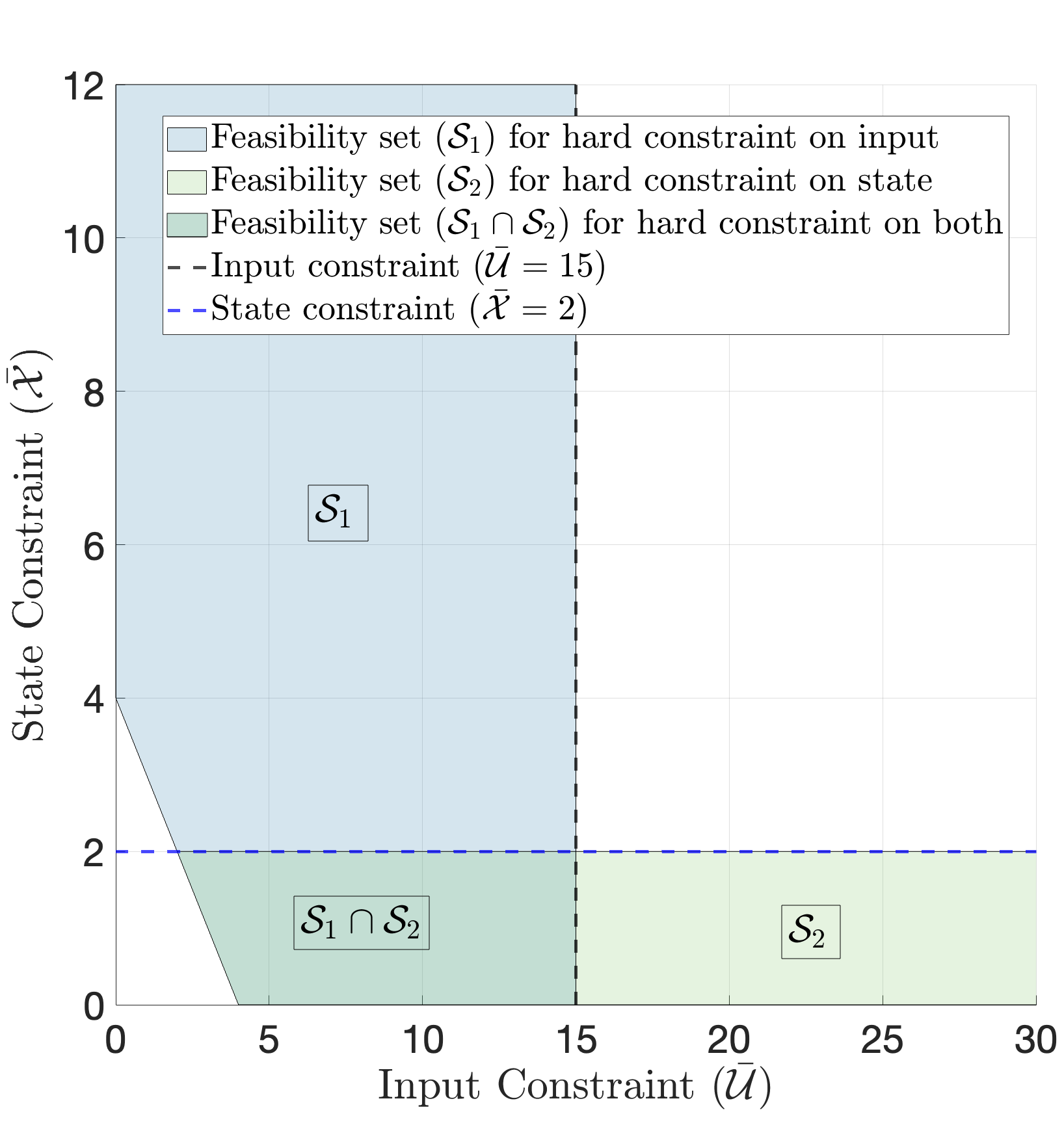}
         \caption{}
         \label{hardcon_input2}
     \end{subfigure}
     \caption{Feasible sets: (a), (b) for varying $\alpha$ with fixed $\beta = 1$; (c) for varying $\beta$ with fixed $\alpha = 1$; (d) for $\alpha > 0$ and $\alpha < 0$; (e), (f) under pre-specified input and/or state bounds for $\alpha > 0$ and $\alpha < 0$, respectively.}
\label{fig1}
     \end{figure}

Feasibility, in this context, refers to the controller's ability to meet the tracking objective while complying with state and input constraints. The feasibility condition \( C1 \) establishes a lower bound on the input constraint, emphasizing the trade-off between the state constraint, the input constraint, and the upper bound of the disturbance. From the feasibility condition C1, we obtain the following inequality.
\begin{align}
\bar{\mathcal{U}}>\alpha\bar{\mathcal{X}}+ \beta \label{f_con}
\end{align}
where 
\begin{align}
    &\alpha =\bar{K}_x-\frac{\lambda_{min}\{Q\}}{2\lambda_{max}\{P\}\|B\|} \label{alpha}\\
    & \beta = \bar{K}_r\bar{r}+\frac{\bar{d}}{\|B\|} +\frac{\bar{\mathcal{X}_a}\lambda_{min}\{Q\}}{2\lambda_{max}\{P\}\|B\|} \label{beta}
\end{align}
From (\ref{abar}) and (\ref{alpha}), it follows that the sign of $\alpha$ indicates the distance between the system matrix \(A\) and the reference model matrix \(A_r\), in the two-norm sense. A larger $\|A_r - A\|$ increases  the required control gain $\bar{K}_x$,  leading to $\alpha > 0$. This implies that if the constraint on the plant states is relaxed, allowing a larger initial tracking error,  the required control effort increases, thereby reducing the feasibility region (Fig. \ref{feasible_alpha}). Conversely, if $\|A_r - A\|$ is sufficiently small such that $\bar{K}_x<\eta$, then $\alpha < 0$. This corresponds to an expanded feasibility region and relaxes the input constraint requirements (Fig. \ref{feasible_negalpha}).
Further, the fixed offset $\beta$ in the feasibility condition captures the minimum control effort required to compensate for the reference input, the external disturbances, and the dynamics of the reference model. Regardless of the sign of $\alpha$, the value of $\beta$ imposes a hard lower bound on $\bar{\mathcal{U}}$. With increasing $\beta$, this bound increases and becomes more restrictive, leading to a reduced feasibility region, as illustrated in Fig. \ref{Hardcon_both}.\\
The evaluation of the controller's feasibility can be approached in the following ways, depending on the user-supplied information regarding state and input constraints. If $\mathcal{\bar{X}}$ or $\mathcal{\bar{U}}$ is not pre-specified by the user, we can analyze feasibility by constructing the region of admissible input-state constraint pairs based on the feasibility condition (\ref{f_con}), parameterized by $\alpha$ and $\beta$. As shown in Fig. \ref{Hardcon_state}, this region includes all $\{\bar{\mathcal{U}}, \bar{\mathcal{X}}\}$ pairs satisfying condition C1. For example, $\{3, 2\}$ is infeasible for $\alpha > 0$, while $\{10, 5\}$ remains feasible irrespective of the sign of $\alpha$. On the other hand, if either of the constraints, input or state, is pre-specified (also denoted as `hard constraint' in Fig. \ref{fig1}), C1 can be used to compute the corresponding admissible bound on the other (represented by $S_1$ or $S_2$), for Cases 2.1 and 2.2, as illustrated in Figs. \ref{hardcon_input} and \ref{hardcon_input2}, respectively. When both constraints are fixed, feasibility is ensured only if the pair lies within the intersection $S_1 \cap S_2$.

\subsection*{Special Case 1: Only state constraint}
In the absence of an input constraint, it follows that $\Delta u(t)=0$, $\forall t\geq0$, and the feasibility condition C1 reduces to the following condition.
\begin{align}
    \bar{\mathcal{X}}>\bar{\mathcal{X}}_a+\frac{2\lambda_{max}\{P\}\bar{d}}{\lambda_{min}\{Q\}}
    \label{onlystate}
\end{align}
For any $\bar{\mathcal{X}}$ that satisfies (\ref{onlystate}), infinite control authority ensures that the plant state remains within the user-defined safe set, i.e. $\|x(t)\|<\bar{\mathcal{X}}$ $\forall t \geq 0$. Further, for the disturbance-free case, no feasibility condition is required, i.e., provided $e(0)\in \Omega_e$, the BLF-based controller ensures the satisfaction of any user-defined state constraint $\bar{\mathcal{X}}>\bar{\mathcal{X}}_r$, however small. 
\subsection*{Special Case 2: Only input constraint}
Unlike Special Case 1, this special case, with only input constraints and no state constraints, cannot be trivially reduced from the feasibility condition C1. In particular, the classical adaptive laws, without BLF modification, can be used
\vspace{-0.15cm}
  \begin{subequations}
  \begin{align}
      &\dot{\hat{K}}_x = \text{proj}_{\Omega_1}\left( -\Gamma_x B^T P e x^T \right)\\
  & \dot{\hat{K}}_r = \text{proj}_{\Omega_2}\left( -\Gamma_r B^T Per^T \right) 
    \end{align} 
\end{subequations}
\vspace{-0.2cm}
along with the reduced feasibility condition 
\begin{align} 
\bar{\mathcal{U}} > \bar{K}_x \bar{\mathcal{X}}_0 + \bar{K}_r \bar{r} + \frac{\bar{d}}{\|B\|} 
\label{sc1}
\end{align} 
where $\bar{\mathcal{X}}_0>0$ is a known auxiliary bound of initial conditions, i.e., $\|x(0)\|<\bar{\mathcal{X}}_0$, guarantees uniform ultimate boundedness (UUB) of the tracking error and boundedness of all closed-loop signals while satisfying the input constraint, i.e., $\|u(t)\|<\bar{\mathcal{U}}$ $\forall t \geq 0$. Detailed Lyapunov analysis is omitted due to space constraints. Note that, when $\bar{K}_x=\eta\implies\sigma=0\implies\alpha=0$, the feasibility condition C1 reduces to $(\ref{sc1})$ with $\bar{\mathcal{X}}_0=\bar{\mathcal{X}}_a$.
\begin{remark}
The feasibility condition C1 is only a sufficient condition for the existence of a feasible control law, i.e., it may be possible to find feasible constraints $\bar{\mathcal{X}}$ and $\bar{\mathcal{U}}$ that do not satisfy C1.
\end{remark}


\section{Simulation Results}
\begin{figure}[htbp]
\centering
\includegraphics[width=\linewidth]{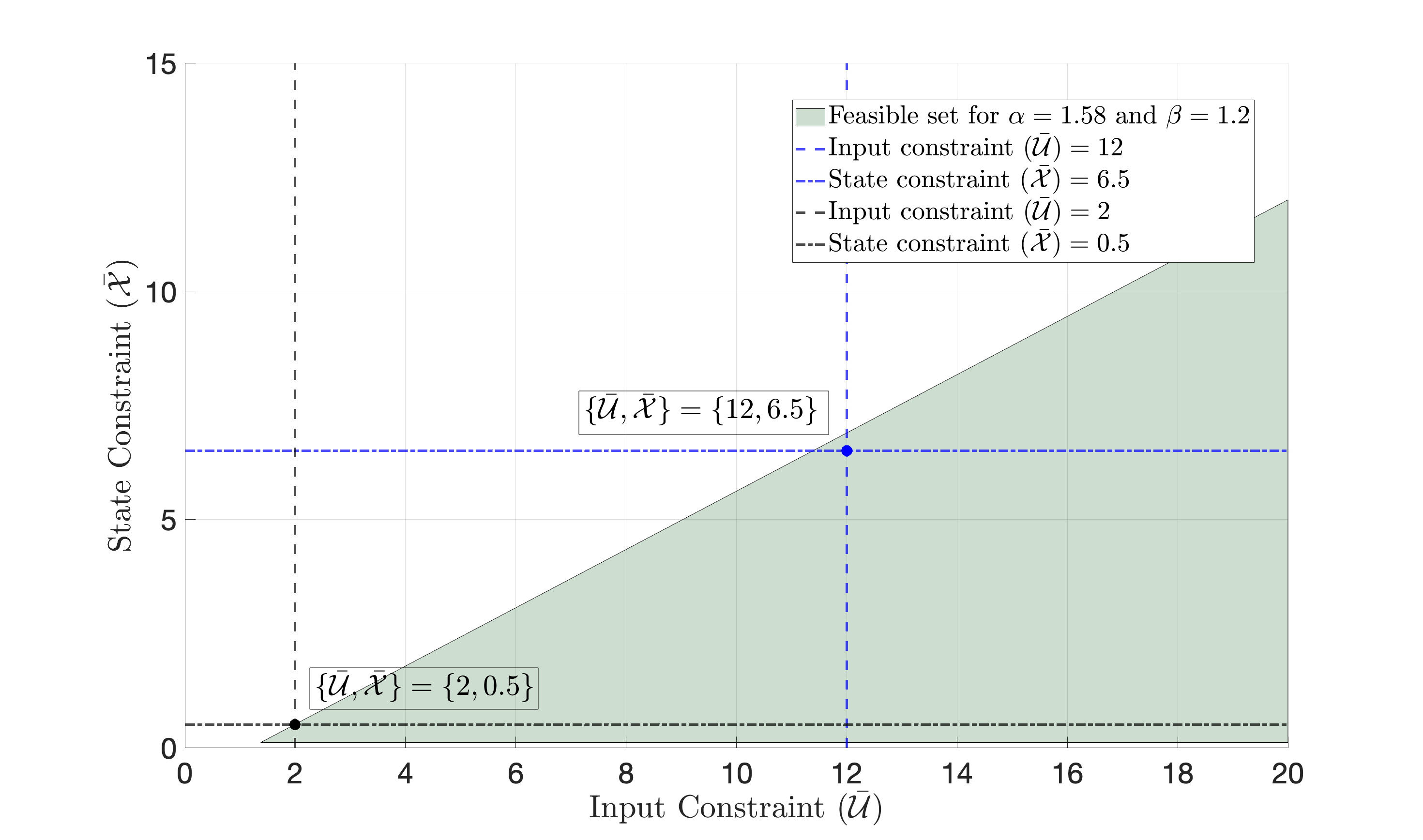}
\caption{Feasibility set for the plant (\ref{plant1}) and the reference model (\ref{ref1}) using (\ref{pc1}), (\ref{pc2}) and (\ref{proposedlaw}).}
\label{feasibility}
\end{figure}

To demonstrate the efficacy of the proposed algorithm, a multivariable unstable LTI plant and a stable reference model are considered.
\begin{equation}
A=\begin{bmatrix}
    0 & 1 & 0 & 0\\
    -1.4 & 1.70 & -0.60 & -0.75\\
    0 & 0 & 0 & 1\\
    -1.5 & -1.50 & -0.95 & -1.5
    \end{bmatrix} 
     \hspace{10pt} B=\begin{bmatrix}
   0 & 0\\
   4 & 0\\
   0 & 0\\
   0 & 3.3
\end{bmatrix} 
\label{plant1}
\end{equation}

\begin{equation}
A_r=\begin{bmatrix}
    0 & 1 & 0 & 0\\
    -5 & -1.5 & -1.2 & -1.6\\
    0 & 0 & 0 & 1\\
    -2 & -3 & -2 & -3
    \end{bmatrix} 
     \hspace{10pt} B_r=\begin{bmatrix}
   0 & 0\\
   1 & 0\\
   0 & 0\\
   0 & 2
\end{bmatrix} 
\label{ref1}
\end{equation}
The objective is for $x(t)$ to track $x_r(t)$ subject to user-defined state and input constraints in the presence of bounded external disturbances. To ensure feasibility the user-defined constraints ($\bar{\mathcal{X}}$ and $\bar{\mathcal{U}}$) will be chosen such that they satisfy C1. Simulation parameters are chosen as follows: $r(t)=[\exp(-t/10);\exp(-t/20)]$, $\bar{\mathcal{X}}_a=6.4$, $Q=\mathbb{I}_4$, $\Gamma_x=5\mathbb{I}_2$, $\Gamma_r=5\mathbb{I}_2$, $\bar{d}=1.2$, $\bar{K}_x=1.6$, $\bar{K}_r=0.6$. Using (\ref{alpha}) and (\ref{beta}), we obtain \(\alpha = 1.58\) and \(\beta = 1.2\) for the given plant and reference model. The feasibility set is illustrated in Fig. \ref{feasibility}.  

\begin{figure}[h!]
\centering
\includegraphics[width=\linewidth]{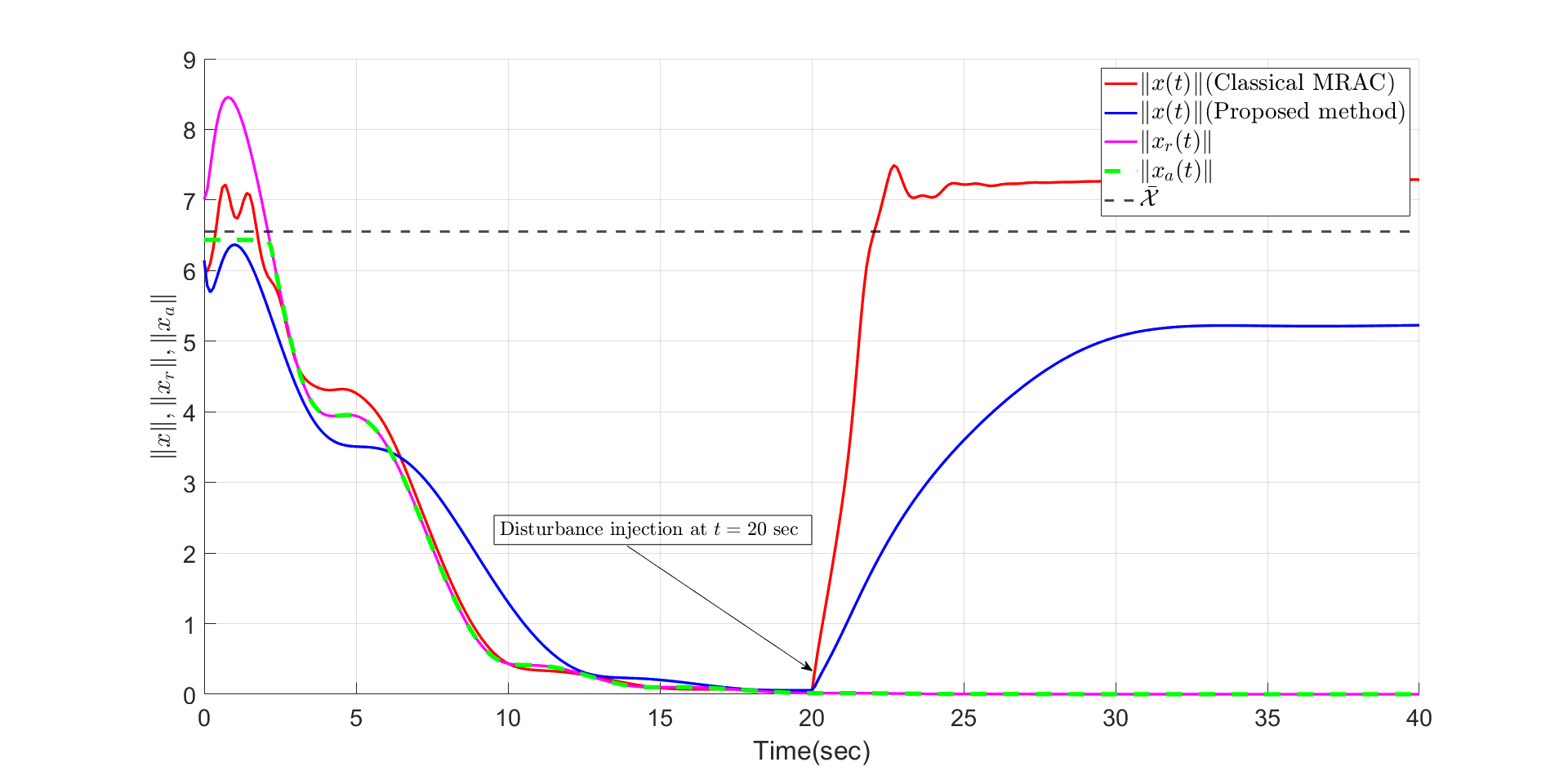}
\caption{Comparative analysis of tracking performance of the plant using proposed controller and robust MRAC.}
\label{tracking_4}
\end{figure}

\begin{figure}[h!]
\centering
\includegraphics[width=\linewidth]{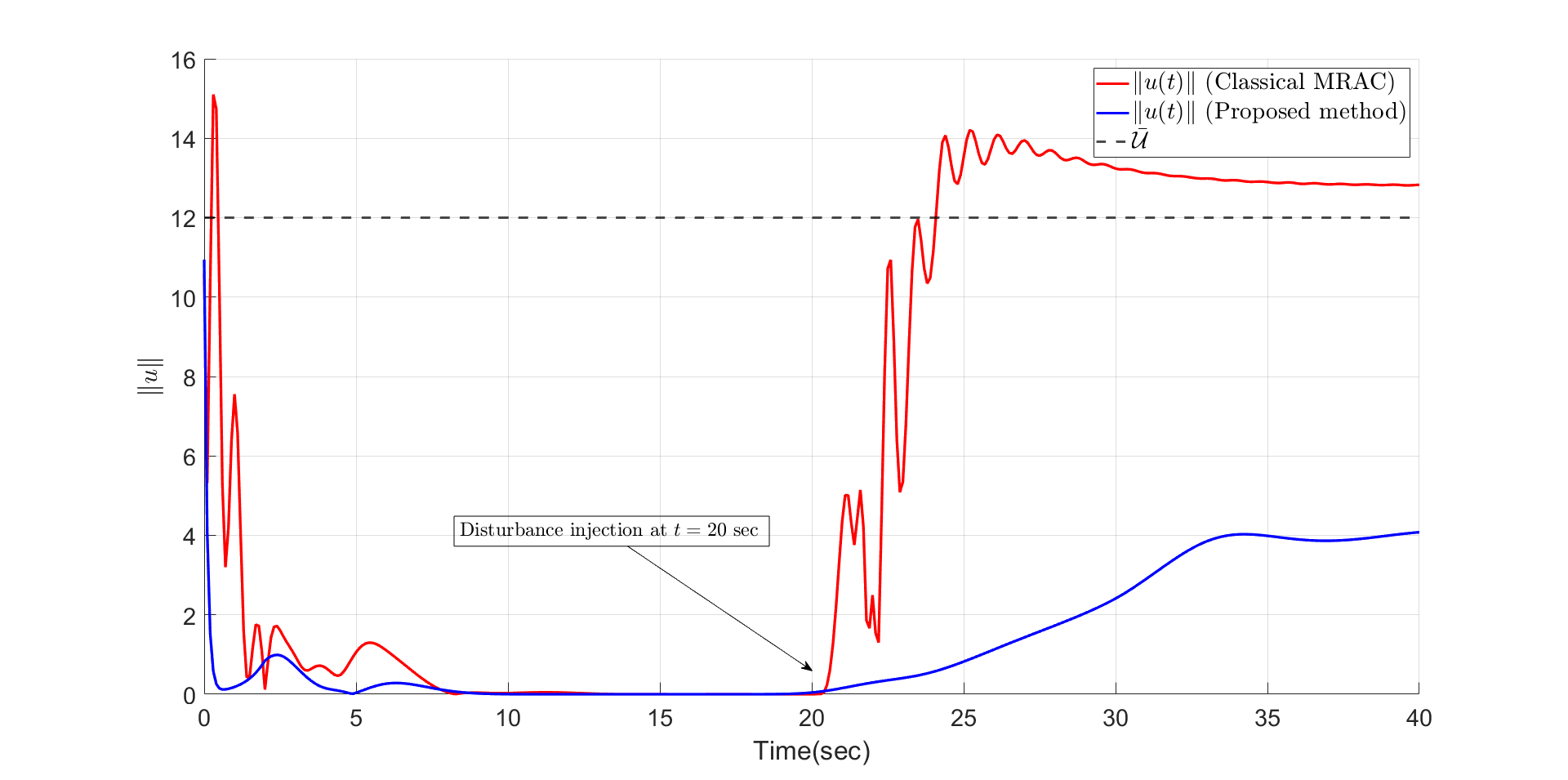}
\caption{Comparative analysis of control input using the proposed controller and robust MRAC.}
\label{input_1}
\end{figure}

To show the efficacy of the proposed control law, we compare it with a standard robust MRAC (using projection) \cite{ioannou1996robust}. The adaptive gains for robust MRAC are chosen as: $\Gamma_x=15\mathbb{I}_2$, $\Gamma_r=15\mathbb{I}_2$. Provided $\bar{d}=1$ and $\bar{\mathcal{X}}=6.5$, the minimum value of the input constraint that satisfies the feasibility condition is calculated as $\bar{\mathcal{U}}=11.5$. For simulation, we have chosen  $\bar{\mathcal{U}}=12$. Fig. \ref{tracking_4} illustrates the tracking performance of both the proposed controller and the robust MRAC. Initially, for $t<20$ sec, there is no external disturbance, and at $t=20$ sec a bounded external disturbance 
with $\bar{d}=1.2$ is introduced in the plant dynamics. The proposed controller consistently satisfies the user-defined state constraint $\forall t \geq 0$, whereas the robust MRAC violates the constraint even in the absence of external disturbance. Additionally, the proposed controller ensures that the required control input remains within the pre-defined safe set (Fig. \ref{input_1}), while for robust MRAC the control input magnitude exceeds the safe region. Note that the adaptation gains for both the proposed controller and robust MRAC are tuned for good tracking performance.
It can be shown that for different initial conditions where $\|x(0)\|<6.5$, the plant states stay within the user-defined bound of $\|x(t)\|<6.5$, even when the initial states are very close to the boundary. Since $\|x_r(0)\|$ is fixed, an increase in $\|x(0)\|$ results in a corresponding rise in $\|e(0)\|$, subsequently leading to larger tracking errors and more control effort.

\begin{remark}
    Higher value of the controller gains, $\Gamma_x$ and $\Gamma_r$ leads to more oscillatory transients and increased control effort but improves steady-state performance and tracking rate. Conversely, reducing the gains smoothens the transient response and lowers the control effort in the transient, but at the expense of slower convergence and reduced steady-state performance. This trade-off must be carefully considered when tuning the controller to meet specific implementation requirements.
\end{remark}

\section{Conclusion}
A constrained MRAC is proposed for LTI systems that accommodates user-defined state and input constraints in the presence of bounded external disturbances. A BLF-based design is strategically merged with a saturated controller to ensure both state and input constraints are satisfied. Adaptive update laws are modified using a projection operator to effectively handle the effect of input saturation and bounded disturbances. Our approach provides an effective alternative to optimization-based constrained control approaches that can be computationally expensive and may become overly conservative for uncertain plants. In this work, we also devise verifiable feasibility criteria to determine the minimum thresholds for the state and input constraints. Under any set of feasible constraints, the proposed controller guarantees that the plant states and the control input remain within the pre-specified safe set while constraining the trajectory tracking error within a known bound. The feasibility conditions demonstrate the relationship between input and state constraints and the associated trade-offs. Simulation results demonstrate the effectiveness of the proposed control law compared to robust MRAC.

\bibliographystyle{ieeetr}
\bibliography{ref}

\end{document}